\documentclass[12pt]{article}
\usepackage{amsfonts}
\title{Basic features of General Superfield Quantization Method
 for gauge theories in Lagrangian formalism}

\author{A.A. Reshetnyak\thanks{E-mail: reshet@tspu.edu.ru}}

\date{\it Seversk State Technological Institute,
 {\rm 636036}, Russia}
\begin{document}

\maketitle
\begin{abstract}
The rules for superfield Lagrangian quantization method for
general gauge theories on a basis of their generalization to
special superfield models within a so-called $\theta$-superfield
theory of fields ($\theta$-STF) are formulated. The
$\theta$-superfield generating functionals of Green's functions
together with effective action are constructed. Their properties
including new interpretation and superfield realization of BRST
transformations, Ward identities are studied.
\end{abstract}

To the construction problem  of the superfield variants of
Lagrangian [1] and canonical [2] quantization methods for gauge
theories realizing the superfield BRST principle one  had been
devoted a number of works [3,4], [5] exploiting trivial and
nontrivial linking of even $t$ and odd $\theta$ supertime $\Gamma
= (t,\theta)$ projections. These investigations are correlated,
in particular,  with appearance of the superfield models
construction methods known as generalized Poisson sigma models
[6] geometrically detailed in [7] and developed further by
Batalin and Marnelius [8].

Lagrangian superfield quantization versions [3,4], [5] possessing
by usual properties of the BV method [1] nevertheless do not
suggest the explicit ways to define in superfield form  the gauge
algebra structural functions determining field theory model that
leads to difficulties when explicit describing of the generating
equations solutions structure. Additionally a specificity   of
generating functional of Green's functions (GFGF)
$Z[\Phi^{\ast}]$ definition [3] appearing in dependence, in
general, of the gauge fermion $\Psi[\Phi]$ and quantum action
${S}[\Phi,{\Phi}^{\ast}]$ on fields $\lambda^A$ (being by
superfield $\Phi^A(\theta)$ components in
$(\Phi^A,{\Phi}_A^{\ast})(\theta)$ = ($\phi^A + \lambda^A\theta,$
$\phi^{\ast}_A - \theta J_A$)) leads to $Z[\Phi^{\ast}]$
difference from BV method GFGF [1]. That fact serves, in
particular, by the obstacle to introduce the superfield effective
action (EA).

By  one from the key ideas in constructing of Lagrangian general
superfield quantization (GSQ) resolving the mentioned
peculiarities of method [3,4] and containing the rules of initial
classical objects representation within $\theta$-STF one have
appeared the enlargement to depend on $\theta$ trivially linking
with $t$ of the all standard gauge field theory objects (action,
GFGF, ...). Intensive use of an analogy with classical Lagrangian
and Hamiltonian mechanics (field theory) realizes the gauge
invariance properties of initial ($\theta$ = 0) model (in
particular, BRST transformations) in the form of Lagrangian and
Hamiltonian systems with dynamical $\theta$. The main results of
the report are presented in [9--11].

By the central and not equivalent Lagrangian $\theta$-STF objects
 are the superfunctional $Z[{\cal A}]$ and $\Lambda_{1}
(\theta;\mathbb{R})$-valued Lagrangian action function
$S_L(\theta)$ = $S_{L}({\cal A} (\theta)$, $\partial_{\theta}{\cal
A}(\theta), \theta)$ defined on odd tangent bundle $T_{odd}{\cal
M}_{cl} \times \{\theta\}$ over configuration space ${\cal
M}_{cl}$ of superfields ${\cal A}^{\imath}(\theta)$ = $A^{\imath}$
+ ${\lambda}^{\imath}\theta$ ($\imath$ = 1,...,$n=n_++n_-$,
$\vec{\varepsilon}({\cal A}^{\imath})$ =
$((\varepsilon_{P})_{\imath}, (\varepsilon_{\bar{J}})_{\imath},
\varepsilon_{\imath})$)
\begin{eqnarray}
Z[{\cal A}] =  \partial_{\theta} S_{L}(\theta), \
\vec{\varepsilon}(Z) = \vec{\varepsilon}(\theta) = (1, 0, 1),
\; \vec{\varepsilon}(S_{L}) = \vec{0},
\end{eqnarray}
with  $Z_2$-gradings $\vec{\varepsilon} = (\varepsilon_{P},
\varepsilon_{\bar{J}}, \varepsilon), (\varepsilon =
\varepsilon_{P} + \varepsilon_{\bar{J}})$ whose auxiliary
components $(\varepsilon_{\bar{J}}$, $\varepsilon_{P})$ are
Grassmann parities w.r.t. superspace ${\cal M}$ = $\tilde{{\cal
M}}\times \tilde{P}$ = $\{(z^M,\theta)\}$ coordinates $z^M$,
$\theta$. ${\cal M}$ may be realized as the
 quotient of global symmetry supergroup  ${J} = \bar{J} \times P$, $P =
 \exp\{\imath\mu p_\theta\}$, $\mu^2$ = $p^2_\theta$ = 0 (for instance,
 with a choice as the  $\bar{J}$ the spacetime SUSY group).

From $T_{\mid \bar{J}}$-scalars $S_L(\theta)$, $Z[{\cal A}]$
w.r.t. action of $J$-superfield  representation $T$ restriction on
$P$: $T_{\mid P}$  the only $S_L(\theta)$ is nontrivially
transformed
 (under transformations ${\cal A}^{\imath}(\theta)$
$\rightarrow$ ${\cal A}'{}^{\imath}(\theta)$ = $(T_{\mid
\bar{J}}{\cal A})^{\imath}(\theta - \mu)$)
\vspace{-1ex}
\begin{eqnarray}
\delta S_L(\theta)  = S_{L}\left({\cal A}'(\theta),
\partial_{\theta}{\cal A}'(\theta), {\theta}\right)
 -S_L(\theta) = - \mu
\left(\textstyle\frac{\partial}{\partial \theta} +
P_0(\theta){\stackrel{\circ}{U}}_+(\theta)\right)S_L(\theta),
\end{eqnarray}
\vspace{-2ex}

\noindent where were introduced the operator $\
{\stackrel{\circ}{U}}_{+}(\theta) =
\partial_{\theta}{\cal A}^{
\imath}(\theta)\textstyle\frac{\partial_{l}
\phantom{xxx}}{\partial {\cal A}^{\imath}(\theta)}$ and projector
$P_{0}(\theta)$ from system  $\{P_a(\theta)$ = $\delta_{a0}(1-
\theta\partial_{\theta})$ + $\delta_{a1}\theta\partial_{\theta}$,
$a=0,1$, $\partial_{\theta} \equiv
\partial^l_{\theta}\}$ acting on $T_{odd}{\cal
M}_{cl} \times \{\theta\}$.

The dynamics of the model is encoded in $\theta$-superfield
Euler-Lagrange equations
\begin{eqnarray}
\frac{\delta_l Z[{\cal A}]}{\delta{\cal
A}^{\imath}(\theta)} = \left(\frac{\partial_l \phantom{xxx}}{
\partial{\cal A}^{\imath}(\theta)}
 -(-1)^{\varepsilon_{\imath}}\partial_{\theta}\frac{\partial_l
\phantom{xxxx}}{\partial(\partial_{\theta}{\cal
A}{}^{\imath}(\theta))}\right)S_{L}(\theta) \equiv
{\cal L}^l_{\imath}(\theta)S_L(\theta) = 0,
\end{eqnarray}
which are equivalent due to identities $\partial_{\theta}^2 {\cal
A}^{\imath}(\theta) = 0$ to  Lagrangian system (LS) of the 2nd
(formally) order on $\theta$ 2$n$ differential equations (DE)
\begin{eqnarray}
&  &
\partial^2_{\theta}{\cal A}^{\jmath}(\theta) \displaystyle\frac{
\partial^{2}_{l} S_{L}(\theta) \phantom{xxxxx}}
{\partial(\partial_{\theta}{\cal
A}^{\imath}(\theta))\partial(\partial_{\theta}{\cal A}^{\jmath}
(\theta))} = \partial^2_{\theta}{\cal
A}^{\jmath}(\theta)(S_{L}'')_{\imath\jmath}(\theta) = 0,
\nonumber \\
 &  & {\Theta}_{\imath}(\theta) = \displaystyle\frac{
\partial_l S_{L}(\theta) }{\partial{\cal A}^{\imath}(\theta)}
 -(-1)^{\varepsilon_{\imath}}\left(\frac{\partial_l}{\partial\theta}\frac{
 \partial_l
S_{L}(\theta)}{\partial(\partial_{\theta}{\cal A}^{\imath}(\theta))} +
{\stackrel{\circ}{U}}_{+}(\theta)\displaystyle\frac{
\partial_l S_{L}(\theta)}
{\partial(\partial_{\theta}{\cal A}^{\imath}(\theta))} \right) =
0.
\end{eqnarray}
Lagrangian differential constraints ${\Theta}_{\imath}(\theta)$ =
${\Theta}_{\imath}({\cal A}(\theta), \partial_{\theta}{\cal
A}(\theta), \theta)$,  firstly, restrict the Cauchy problem
setting for  LS and, secondly, may be as the 1st order on $\theta$
DE  by functionally dependent. Provided that a  local supersurface
$\Sigma$ exists the such that
\begin{eqnarray}
{\Theta}_{\imath}(\theta)_{\textstyle\vert\Sigma}\hspace{-0.2em}=0,
\dim\Sigma = (m_+,m_-),\; {\rm
rank}_{\varepsilon}\left\|\hspace{-0.1em}\partial_{\theta}\frac{\delta_l
\phantom {xxxx}}{\delta{\cal A}^{\jmath}(\theta_1)} \frac{\delta_l
Z[{\cal A}]}{\delta{\cal
A}^{\imath}(\theta)}\hspace{-0.1em}\right\|_{\textstyle\vert\Sigma} \hspace{-0.5em} = (n_+,n_-)-(m_+,m_-),  
\end{eqnarray}
\vspace{-2ex}

\noindent there are at least $m=m_++m_-$ differential identities
(for $\alpha_0=1,...,m_0=m_{0+}+m_{0-}$)
\begin{eqnarray}
\int\hspace{-0.3em} d\theta \frac{\delta_r Z[{\cal A} ]}{\delta
{\cal A}^{\imath}(\theta)}{}{\hat{{\cal R}}}^{\imath}_{\alpha_0}
(\theta; {\theta}')\hspace{-0.1em} = \hspace{-0.1em}0,\
{\hat{{\cal R}}}^{\imath}_{\alpha_0}(\theta;
{\theta}')\hspace{-0.1em} = \hspace{-0.1em}\sum_{k\geq 0}
\left(\left(\partial_{\theta} \right)^k \delta(\theta -
\theta')\right) {\hat{ {\cal R}}}_{k}{}^{\imath}_{\alpha_0} \bigl(
{\cal A}(\theta),
\partial_{\theta}{\cal A}(\theta), \theta\bigr),
\end{eqnarray}
 with  functionally dependent for $(m_{0+},m_{0-})>(m_+,m_-)$
generators ${\hat{\cal R}}^{\imath}_{\alpha_0}\bigl({\cal
A}(\theta),
\partial_{\theta}{\cal A}(\theta), \theta;$ ${\theta}'\bigr)$ of
general  gauge transformations: $\delta_{g}{\cal
A}^{\imath}(\theta)$ = $\int\hspace{-0.1em} d\theta' {\hat{{\cal
R}}}^{\imath}_{\alpha_0} (\theta;
{\theta}'){\xi}^{\alpha_0}(\theta')$, leaving $Z[{\cal A}]$ by
invariant. In the case  of the $\theta$-STF model representability
in the natural system form: ($S_L ( \theta)$ = $T
\bigl(\partial_{\theta}{\cal A}(\theta)\bigr)$ $-$ $S\bigl( {\cal
A}(\theta), \theta\bigr))$, functions ${\Theta}_{\imath}(\theta)$
pass into holonomic constraints  given on ${\cal M}_{cl}$ $\times$
$\{\theta\}$ (for $\theta$ = 0 being by  usual extremals for
${\cal S}_0(A)$ = $S\bigl( {\cal A}(0), 0\bigr)$)
\begin{eqnarray}
{\Theta}_{\imath}(\theta) = - S,_{\imath} \bigl( {{\cal
A}}(\theta), \theta \bigr)(-1)^{ \varepsilon_{\imath}} = 0,\
S,_{\imath}(\theta) =
\frac{\partial_r S(\theta)}{\partial{\cal A}^{\imath}(\theta)}.   
\end{eqnarray}
 Condition (5), identities (6) take the (usual for
$\theta=0$) form
\begin{eqnarray}
{\rm rank}_{\varepsilon}\left\|S,_{\imath \jmath} \bigl( {{\cal
A}}(\theta),\theta)\right\|_{ \textstyle \mid \Sigma} =
(n_+,n_-)-(m_+,m_-) ,\ S,_{\imath}\bigl({\cal A}(\theta), \theta
\bigr) {{\cal R}}_{0}{}^{\imath}_{\alpha_0}\bigl( {\cal
A}(\theta), \theta\bigr) = 0,
\end{eqnarray}
with linearly dependent  generators of special gauge
transformations (SGT): $\delta{\cal A}^{\imath}(\theta)$ = ${\cal
R}_{0}{}^{\imath}_{\alpha_0} \bigl({\cal A}(\theta), \theta
\bigr){\xi}_{0}^{\alpha_0}(\theta)$ w.r.t. which the only
$S(\theta)$  is invariant. Relations (7, 8) define  the reducible
special type gauge theory (GThST) in contrast to one of general
type (GThGT), $\theta$-locally including GThST, with more general
expressions for $S_L(\theta)$, $\Theta_{\imath}(\theta)$.

Independently $\theta$-STF model is formulated without ${\cal
M}_{cl}$ extension
 in terms of Hamiltonian classical action superfunction $S_H(\Gamma(\theta),
 \theta)$  given on $T^{\ast}_{odd}{\cal M}_{cl}
\times\{\theta\}$ = $\{(\Gamma^p(\theta), \theta)\}$ with
superantifields (${\cal A}^{\ast}_{\imath}(\theta)$ =
$A^{\ast}_{\imath}$ $-$ $\theta J_{\imath})$ included in
$\Gamma^p(\theta)$ = $({\cal A}^{\imath}, {\cal
A}^{\ast}_{\imath})(\theta)$, ($\vec{\varepsilon}({\cal
A}^{\ast}_{\imath})$ = $\vec{\varepsilon}({\cal A}^{\imath})$ +
$(1,0,1))$. In that case the equivalence of the  Hamiltonian and
Lagrangian $\theta$-STF model formulations is ensured by  the
supermatrix  $\left\|(S_{L}'')_{\imath\jmath}(\theta)\right\|$
 nondegeneracy within Legendre transform of
$S_L(\theta)$ w.r.t. $\partial_{\theta}{\cal A}^{\imath}( \theta)$
\begin{eqnarray}
S_H(\Gamma(\theta), \theta) = {\cal A}^{\ast}_{\imath}(
\theta)\partial^r_{\theta}{\cal A}^{\imath}(\theta) - S_{L}(\theta
),\; {\cal A}^{\ast}_{\imath}(\theta) = \frac{\partial_r
S_L(\theta)\phantom{xx}}{\partial(\partial^r_{\theta} {\cal
A}^{\imath}(\theta))}. 
\end{eqnarray}
The generalized Hamiltonian system  [10] as the 1st order on
$\theta$ 3n DE being equivalent to LS reads in terms of
$\theta$-local antibracket as follows
\begin{eqnarray}
\partial^r_{\theta}\Gamma^p(\theta) =
\left(\Gamma^p(\theta), S_H(\theta)\right)_{\theta}, \
\Theta^H_{\imath}(\Gamma(\theta), \theta) = \Theta_{\imath}({\cal
A}(\theta), \partial_{\theta}{\cal A}( \Gamma(\theta),\theta),
\theta)  = 0, 
\end{eqnarray}
with  generalized Hamiltonian constraints
$\Theta^H_{\imath}(\Gamma(\theta), \theta)$ coinciding with half
equations from  properly HS due to (9) and their consequences
\begin{eqnarray}
\Theta^H_{\imath}(\Gamma(\theta), \theta) = -
\left(\partial^r_{\theta}{\cal A}^{\ast}_{\imath}(\theta) +
S_H,_{\imath}(\theta)(-1)^{\varepsilon_{\imath}}\right),
\end{eqnarray}
that establishes  the equivalence of HS with generalized HS and
therefore with LS jointly with corresponding Cauchy problems
setting for $\theta=0$ [10] determining  integral curves
$\overline{\Gamma}{}^p(\theta), \bar{\cal A}^{\imath}(\theta)$.
In order to superfunction \vspace{-2ex}
\begin{eqnarray}
S_{E}\bigl({\cal A}(\theta),
\partial_{\theta}{\cal A}(\theta),
\theta\bigr) = \frac{\partial_r
S_L(\theta)\phantom{xx}}{\partial(\partial^r_{\theta} {\cal
A}^{\imath}(\theta))}\partial^r_{\theta}{\cal A}^{\imath}(\theta)
- S_{L}(\theta), 
\end{eqnarray}
obtained by the Noether's Theorem 1 application for the
translation along $\theta$ on  constant $\mu$, as the symmetry
transformation  for $(d\theta S_L(\theta))$,  would be by LS
integral (i.e. by the quantity being conserved  w.r.t.
$\theta$-evolution) it is sufficient to fulfill the equations
\begin{eqnarray}
\textstyle\frac{\partial}{
\partial \theta}
S_L(\theta)=0, \
\left(S_{L},_{\imath}(\theta)\partial^r_{\theta}{\cal
A}^{\imath}(\theta)\right)_{
\vert {\cal L}_{\imath}^{l}S_{L} = 0}=0. 
\end{eqnarray}
In Hamiltonian formulation Eqs.(13) are reduced to the expressions
meaning the invariance of $S_H(\Gamma(\theta),\theta)$ =
$S_{E}\bigl({\cal A}(\theta)$, $\partial_{\theta}{\cal
A}(\Gamma(\theta), \theta)$, $\theta\bigr)$ under
$\theta$-translations along $\overline{\Gamma}{}^p(\theta)$
\begin{eqnarray}
   &  \delta_\mu S_H(\theta)_{\vert
\overline{\Gamma}(\theta)}  = \mu  \left[\textstyle\frac{\partial
}{\partial \theta}S_H(\theta) -  \left(S_H(\theta), S_H(
\theta)\right)_{\theta}\right] =0 \Rightarrow &\\
  & \textstyle\frac{\partial }{\partial \theta}S_H(\theta) =0,\
{\left(S_H(\theta), S_H(\theta)
\right)_{\theta}}=0. &
\end{eqnarray}
Eqs.(15) fulfillment jointly with $\theta$-superfield
integrability for HS (10) is provided by the presence of special
Hamiltonian constraints  $(\frac{\partial
S_H(\Gamma(\theta))}{\partial {\cal A}^{\ast}_{
A_1}(\theta)\phantom{x}}$, $\frac{\partial_l
S_H(\Gamma(\theta))}{\partial {\cal A}^{
A_2}(\theta)\phantom{xx}}) = 0$, $(A_1 \cup A_2)$ = $\imath$.
Really in view of Jacobi identity for antibracket the latter
property is based on the relationship
\begin{eqnarray}
0=\partial^2_{\theta}\Gamma^p(\theta) = - {\textstyle\frac{1}{2}
\bigl(\Gamma^p(\theta), \bigl(S_H(\Gamma(\theta)),
S_H(\Gamma(\theta)) \bigr)_{
\theta}\bigr)_{\theta}} = 0, 
\end{eqnarray}
so that the translation formula on superalgebra
$C^k(T^{\ast}_{odd}{\cal M}_{cl}\times\{\theta\})$, $k\leq
\infty$ holds
\begin{eqnarray}
\delta_{\mu}{\cal F}(\theta)_{\vert\overline{\Gamma}(\theta)}
\hspace{-0.15em} =  \hspace{-0.15em} \mu
\partial_{\theta}^l{\cal F}(\theta)_{\vert\overline{\Gamma}(\theta)}
\hspace{-0.15em} = \hspace{-0.15em} \mu
\left(\hspace{-0.15em}\textstyle\frac{\partial^l}{\partial
\theta\phantom{}}\hspace{-0.1em} -\hspace{-0.1em} {\rm
ad}{}S_H(\theta)\right){\cal F}(\theta),\; {\rm ad}{}S_H(\theta)=
(S_H(\theta),\ )_{\theta}.
\end{eqnarray}
The HS solvability and nilpotency of the BRST type generator of
$\theta$-shifts along solutions $\check{s}^l(\theta)$ =
$\frac{\partial}{\partial\theta} - {\rm ad}{}S_H(\theta)$ are
guaranteed by the master-equation fulfillment in
$T^{\ast}_{odd}{\cal M}_{cl}$   additionally to which one may be
supposed the equation validity $\bigr((\Gamma^d(\theta),
\Gamma^q(\theta))_{\theta}\omega_{qp}(\theta)$\nolinebreak{\,}=\nolinebreak{
\,}$\delta^d{}_p\bigr)$
\begin{eqnarray}
\Delta^{cl}(\theta)S_H(\Gamma(\theta)) = 0,\; \Delta^{cl}(\theta)
=  \textstyle\frac{1}{2}
(-1)^{\varepsilon(\Gamma^q)}\omega_{qp}(\theta)\bigl(\Gamma^p(\theta),
\bigl(\Gamma^q(\theta),\
\ \bigr)_{\theta}\bigr)_{\theta}. 
\end{eqnarray}
Functions $\Theta^H_{\imath}(\Gamma(\theta), \theta)$ permit  the
analogous to $\Theta_{\imath}(\theta)$ (4) analysis of one's
functional dependence with corresponding interpretation for gauge
functions and relations.

The quantization within GSQ consists in the restriction of GThGT
in any $\theta$-STF formulations   by the prescriptions $({\rm
gh},\textstyle\frac{\partial\phantom{x}}{\partial\theta})
S_{H(L)}(\theta)$  = $(0,0)$, with  standard ghost number
distribution for $\Gamma^p_{cl}(\theta)$ [1]. Solution for these
equations  in assuming of the potential term existence in
$S_{L(H)}(\theta)$: $ S({\cal A}(\theta),0)$ = ${\cal S}_0({\cal
A}(\theta))$, transfers generalized HS (10) into $\theta$-solvable
system with holonomic $\Theta_{\imath}^{H}({\cal A}(\theta))$ =
$\Theta_{\imath}( {\cal A}(\theta))$
\begin{eqnarray}
\{\partial^r_{\theta}{\cal A}^{\imath}(\theta),
\partial^r_{\theta}{\cal A}^{\ast}_{\imath}(\theta),
\Theta_{\imath}^{H}({\cal A}(\theta))\} = \{0, -
(-1)^{\varepsilon_{\imath}}{\cal S}_{0},_{\imath}({\cal
A}(\theta)), - (-1)^{\varepsilon_{\imath}}{\cal
S}_{0},_{\imath}({\cal A}(\theta))\}. 
\end{eqnarray}
The substitution  $(\xi_0^{\alpha_0}(\theta)$ = $
d\tilde{\xi}^{\alpha_0}_{0}(\theta)$ =
$C^{\alpha_0}(\theta)d\theta)$ permits to extend SGT to
Hamiltonian system with $2n$ DE built w.r.t. function
$S_1((\Gamma_{cl},C)(\theta))$ = $ {\cal
A}^{\ast}_{\imath}(\theta){\cal R}_{0}{}^{\imath}_{\alpha_0}({\cal
A }(\theta))C^{\alpha_0}(\theta)$  the enlargement of the union of
which with Eqs.(19) up to $2(n + m)$ DE has the form of HS
\begin{eqnarray}
\partial^r_{\theta}\Gamma^p_{s}(\theta) = \left(\Gamma^p_{s}(\theta),
S_{[1]}(\theta)\right)^{s}_{\theta},\ S_{[1]}(\theta) = {\cal
S}_0({\cal A}(\theta))
+ S_{1}(\theta), s=min. 
\end{eqnarray}
For reducible theory the sets $\Gamma^p_{min}(\theta)$ =
$(\Gamma_{cl}, C^{\alpha_0}, C^{\ast}_{\alpha_0})(\theta)$,
$S_{[1]}(\theta)$, HS (20) are extended by means of the ghost
supervariables pyramid parameterizing ${\cal M}_s$ =
$\{\Phi^A_s(\theta)\}$, $T^{\ast}_{odd}{\cal M}_s$ = $\{(\Phi^A_s,
\Phi^{\ast}_{A{}s})(\theta)\}$. $\theta$-superfield integrability
for  (20) is provided by the function $S_{[1]}(\theta)$
deformation in powers of ghost superfields guaranteeing within
theorem existence [11] the obtaining of proper [1,11] solution of
the classical master-equation in minimal sector
\begin{eqnarray}
 (S_{H{}s}(\Gamma_{s}(\theta)),S_{H{}s}(\Gamma_{s}(\theta)))^s_{\theta}=
0,\ (\vec{\varepsilon}, {\rm gh},
\textstyle\frac{\partial}{\partial\theta})S_{H{}s}(\Gamma_{s}(\theta))=
(\vec{0}, 0,0)  
\end{eqnarray}
in the case of general (open) gauge algebra of ${\cal
R}_{0}{}^{\imath}_{\alpha_0}({\cal A }(\theta))$.

Enlargement of $S_{H{}min}(\theta)$ until Eqs.(21) proper solution
in $T^{\ast}_{odd}{\cal M}_{s}$ = $\{\Gamma_{s}(\theta)\}$,
$\Gamma_{s}(\theta)$ = $(\Gamma_{min}, \overline{C}{}^{\alpha_0},
B^{\alpha_0}$, $\overline{C}{}^{\ast}_{\alpha_0},
B^{\ast}_{\alpha_0})(\theta)$ (in what follows $s = ext$) and
deformation in $\hbar$ powers ensures the quantum action
$S_{H}^{\Psi}(\Gamma_{s}(\theta), \hbar)$ construction for abelian
hypergauge determined by phase anticanonical transformation
\begin{eqnarray}
\Gamma^p_s(\theta) \rightarrow {\Gamma'}^p_s(\theta) =
\left(\Phi^A(\theta), {\Phi}^{\ast}_A(\theta) -
\frac{\partial\Psi(\Phi(\theta))}{
\partial\Phi^A(\theta)\phantom{x}}
\right): S_{H}^{\Psi}(\Gamma_{s}(\theta), \hbar) = e^{{\rm
ad}_{\Psi}} S_{H{}s}(\Gamma_{s}(\theta),\hbar). 
\end{eqnarray}
$S_{H}^{\Psi}(\Gamma(\theta), \hbar)$ just as
$S_{H{}s}(\theta,\hbar)$ satisfies to Eqs.(18, 21)  if
$S_{H{}min}(\theta,\hbar)$ had been the same.

Representation  $S_{H{}s}(\theta,\hbar)$ in the forms (in omitting
of $\theta$-dependence sign)
\begin{eqnarray}
 \tilde{S}_{H}(\hbar) = S_{H{}min}(\hbar) +
\overline{C}{}^{\ast}_{\alpha_0}B^{\alpha_0},\ {S}_{H{}s}(\hbar)
= S_{1{} s}\bigl((\Gamma,
\partial_{\theta}{\Gamma}),\hbar\bigr)
+(\partial_{\theta}{\Phi}^{\ast}_{A})\Phi^{A} 
\end{eqnarray}
identifies (for irreducible GThST)
$P_0(\theta)\tilde{S}_{H}(\theta,\hbar)$ with
$S_{BV}(\Gamma_{s}(0),\hbar)$ [1] satisfying to (21) with
antibracket $(\ ,\ )_{BV}$, calculated w.r.t. $\Gamma_{s}(0)$ =
$\Gamma_{0s}$, and $S_{1{} s}(\theta,\hbar)$ (co\-in\-ci\-ding for
$\theta$ = 0 with ${S}[\Phi,{\Phi}^{\ast}]$ [3]) obeying to
$\theta$-su\-per\-fi\-e\-ld generating equation [3] with
$\Delta^s(\theta)S_{1{} s}(\theta,\hbar) = 0$
\begin{eqnarray}
\textstyle\frac{1}{2}\Bigl(S_{1{}s}(\theta,\hbar),
S_{1{}s}(\theta,\hbar)\Bigr)_{\theta} +
{\stackrel{\circ}{V}}_{+}(\theta)S_{1{}s}(\theta,\hbar) = 0,\
{\stackrel{\circ}{V}}_{+}(\theta) =
\partial_{\theta}\Phi^{\ast}_A(\theta)
\displaystyle\frac{\partial\phantom{xxx}}{\partial
\Phi^{\ast}_A(\theta)
}. 
\end{eqnarray}
Solvable HS built w.r.t. $S_{H}^{\Psi}(\theta, \hbar)$ and
non-equivalent to HS with Hamiltonian $S_{H{} s}(\Gamma_s(\theta),
\hbar)$ plays for $\theta$ = 0 the key role in [1] defining the
$\theta$-superfield (not nilpotent) BRST transformations
generator  annulling $S_{H}^{\Psi}(\theta, \hbar)$ [11] and
associated with its nontrivial subsystem
\begin{eqnarray}
\partial_{\theta}^r{\Phi}^{A}(\theta) =
\bigr(\Phi^A(\theta),
S^{\Psi}_{H}(\theta,\hbar)\bigr)^{(\Gamma_{s})}_{ \theta},\
\tilde{s}_l^{\Psi}(\theta) = \frac{\partial}{\partial\theta} +
\frac{\partial_r S^{\Psi}_{H}(\theta,\hbar)}{\partial
\Phi^{\ast}_A(\theta)\phantom{xxx}} \frac{\partial_l
\phantom{xxx}}{\partial\Phi^A(\theta)}. 
\end{eqnarray}
GFGF $Z\bigl(\Phi^{\ast},
\partial_{\theta}\Phi^{\ast}\bigr)(\theta)$ as the function
of sources $\partial_{\theta}\Phi^{\ast}_A(\theta)$ = $- J_{A}$,
$(({\rm gh}, \vec{\varepsilon})\partial_{\theta}\Phi^{\ast}_A$ =
$(-{\rm gh}$, $\vec{\varepsilon})\Phi^A)$ is defined, for instance
[11], by means of path integral of functions for fixed $\theta$
(with usual within perturbation theory properties and measure
$d\Phi(\theta)$ = $\prod_A d\Phi^A(\theta))$
\begin{eqnarray}
Z\bigl(\Phi^{\ast}, \partial_{\theta}\Phi^{\ast}\bigr)(\theta) =
 \int
d\Phi(\theta) \exp\left\{\frac{\imath}{\hbar}\left(
{S}^{\Psi}_{H}\bigl({\Gamma}(\theta),\hbar\bigr) -
((\partial_{\theta}\Phi^{\ast}_A)\Phi^A(\theta))\right)\right\}. 
\end{eqnarray}
The actions ${S}^{\Psi}_{H}(\theta,\hbar)$, ${S}_{H{}s}
(\theta,\hbar)$ deformation in powers of
$\partial_{\theta}\Gamma^p(\theta)$ subject to Eqs.(18, 21)
validity (now not anticanonically invariant) permits to construct
$ Z\bigl(\Phi^{\ast}, \partial_{\theta}\Phi^{\ast}$,
$\partial_{\theta}\Gamma^p\bigr)(\theta)$ as GFGF parametrically
depending on $\partial_{\theta}\Gamma^p$ [11], in first, providing
the connection for $\theta=0$  with  $Z[\Phi^{\ast}]$ [3] found in
[11]. Secondly, that fact leads to the $\theta$-superfield
extension of GFGF $Z((\Phi^{\ast},
\partial_{\theta}\Phi^{\ast})(0))$ [1] in
$\partial_{\theta}\Gamma^p$ powers compatible with the way of
gauge imposing in BV method. Therefore EA
{\boldmath$\Gamma$}(($\langle\Phi\rangle, \Phi^{\ast})(\theta))$ =
{\boldmath$\Gamma$}($\theta$)
\begin{eqnarray}
{\mbox{\boldmath$\Gamma$}}\bigl(\langle\Phi\rangle,
\Phi^{*}\bigr)(\theta) = \frac{\hbar}{\imath} \ln
Z\bigl(\Phi^{\ast},
\partial_{\theta}\Phi^{\ast}\bigr)(\theta) +
((\partial_{\theta}\Phi^{\ast}_A)\langle\Phi^A\rangle)(\theta),\
\langle\Phi^{A}(\theta)\rangle = -\frac{\hbar}{\imath}
\frac{\partial_l \ln Z(\theta)\phantom{}}{\partial(\partial_{\theta}\Phi^{\ast}_A(
\theta))}, 
\end{eqnarray}
is given for parametric dependence w.r.t.
$\partial_{\theta}\Gamma^p$ resolving the mentioned peculiarity of
method [3]. HS built w.r.t. $S_{H}^{\Psi}(\theta,\hbar)$ permits
to obtain the usual [1] and new properties for $Z(\theta)$.
Corresponding to this HS with solution $\check{\Gamma}(\theta)$
change  $\Gamma(\theta)$ $\to$ $\Gamma^{(1)}(\Gamma(\theta))$
\begin{eqnarray}
\Gamma^{(1){}p}(\theta) = \exp\{\mu s_l^{\Psi}(\theta)\}
\Gamma^p(\theta),\ \ s_l^{\Psi}(\theta) =
\textstyle\frac{\partial}{\partial \theta} - {\rm
ad}{}S_{H}^{\Psi}(\theta,\hbar)  
\end{eqnarray}
is anticanonical transformation with ${\rm
Ber}\|\frac{\partial\Gamma^{(1)}(\theta)}{\partial\Gamma(\theta)
\phantom{x}}\|$ = ${\rm
Ber}\|\frac{\partial\Phi^{(1)}(\theta)}{\partial\Phi(\theta)
\phantom{x}}\|$ = 1.

Vacuum function $Z_{\Psi}(\Phi^{\ast}(\theta))$ =
$Z(\Phi^{\ast}(\theta),0)$ is invariant both w.r.t. anticanonical
transformation  (28) (formally) and w.r.t. the same change of
variables \vspace{-1ex}
\begin{eqnarray}
Z_{\Psi}^{(1)}(\Phi^{(1){}\ast}(\theta)) & = & \int
 d\Phi^{(1)}(\theta)
\exp\left\{\frac{\imath}{\hbar}
S_H^{(1){}\Psi}(\Gamma^{(1)}(\theta),\hbar)\right\}=
Z_{\Psi}(\Phi^{\ast}(\theta)), \nonumber \\
Z_{\Psi}(\Phi^{(1){}\ast}(\theta)) & = &  \int
d\Phi^{(1)}(\theta) \exp\left\{\frac{\imath}{\hbar}
S_H^{\Psi}(\Gamma^{(1)}(\theta),\hbar)\right\}=
Z_{\Psi}(\Phi^{\ast}(\theta)), 
\end{eqnarray}
\vspace{-2ex}

\noindent and does not depend upon small variation of the gauge
fermion $\Psi(\Phi(\theta))$. Eqs.(18, 21) for
$S_{H}^{\Psi}(\theta,\hbar)$ allow to reproduce  for $Z(\theta)$,
{\boldmath$\Gamma$}$(\theta)$ (the standard  for $\theta=0$) Ward
identities
\begin{eqnarray}
{\stackrel{\circ}{V}}_{+}(\theta)Z(\theta)=0,\ \
\bigl(\mbox{\boldmath$\Gamma$}(\theta),
\mbox{\boldmath$\Gamma$}(\theta)\bigr)^{(\langle\Gamma\rangle)}_{
\theta}=0, 
\end{eqnarray}
\vspace{-2ex}

\noindent and with regards for permutability rule:
$\partial_{\theta}$ $\int d\Phi(\theta) = \int
d\Phi(\theta)(\frac{\partial}{\partial{\theta}} +
{\stackrel{\circ}{V}}_{+}(\theta))$, to obtain the relations
\begin{eqnarray}
\partial_{\theta}Z(\theta)_{\vert\check{\Gamma}(\theta)} =
{\stackrel{\circ}{V}}_{+}(\theta)Z(\theta)=0,\ \
\partial^r_{\theta}{\mbox{\boldmath$\Gamma$}}(\theta)_{\vert
\check{\Gamma}(\theta)} = \bigl(\mbox{\boldmath$\Gamma$}(\theta),
\mbox{\boldmath$\Gamma$}(\theta)\bigr)^{(\langle\Gamma\rangle)}_{\theta}=0
, 
\end{eqnarray}
revealing the fact of GFGF invariance  under the
$\theta$-superfield BRST type transformations generated by HS in
question. In deducing of the Eqs.(31) it is implied the use of
corresponding average of the HS w.r.t. $Z(\theta)$,
{\boldmath$\Gamma$}$(\theta)$.

Note the GSQ construction may be generalized both to the case of
curved  ${\cal M}_s$, $T^{\ast}_{odd}{\cal M}_s$ and for
nonabelian hypergauges. Secondly, the standard BRST
transformations definition in BV method from superfield ones is
given by the Eqs.(25) with  solution $\tilde{\Gamma}(\theta)$
\begin{eqnarray}
& & P_0
\delta_{\mu}\Gamma^{p}(\theta)_{\vert\tilde{\Gamma}(\theta)} =
P_0 \left(\partial^r_{\theta}\Gamma^{p}(\theta)\right)_{
\vert\tilde{\Gamma}(\theta)}\mu = \delta_{\mu}{\Gamma}^{p}_{0},
\nonumber \\
& & \delta_{\mu}\phi^A = \left(\phi^A,
{S}_H\left(\phi,\phi^{\ast} +
\frac{\delta{\Psi}}{\delta\phi\phantom{x}},\hbar\right)
\right)_{BV}\mu,\ \ \delta_{\mu}\phi^{\ast}_A = 0.
\end{eqnarray}
Thirdly, the $\theta$-component  formulation of the superfield
objects and operations from $\theta$-STF [9,10] jointly with
connection establishment of the GSQ quantities with ones from BV
[1] and superfield [3] methods  are detailed in [11].
\begin{center}
{\large{\bf References}}
\end{center}
\begin{enumerate}
\item I.A. Batalin, G.A. Vilkovisky,
Phys. Lett. B102 (1981) 27; Phys.  Rev. D28 (1983) 2567.
\item E.S. Fradkin,  G.A. Vilkovisky, Phys. Lett. B55 (1975)
224;\\
I.A. Batalin, G.A. Vilkovisky, Phys. Lett. B69 (1977) 309;\\
E.S.Fradkin, T.E. Fradkina, Phys. Lett. B72 (1978) 343;\\
I.A. Batalin, E.S. Fradkin, Phys. Lett. B122 (1983) 157.
\item P.M. Lavrov, P.Yu. Moshin,
A.A. Reshetnyak, Mod. Phys. Lett. A10 (1995) 2687; JETP Lett. 62
(1995) 780.
\item B. Geyer, P.M. Lavrov, P.Yu. Moshin, Phys. Lett. B463 (1999) 188;\\
P.M. Lavrov, P.Yu. Moshin, Superfield Covariant Quantization with
BRST Symmetry, in: Proc. of the Int. Conf. dedicated to Memory of
E. Fradkin: Quantization, Gauge Theories and Strings, ed. A.
Semikhatov, M. Vasiliev, V.Zaikin, Scientific World, 2 (2001) 205.
\item I.A. Batalin, K. Bering,  P.H. Damgaard,
Nucl. Phys. B515 (1998) 455; Phys. Lett. B446 (1999) 175.
\item A.S. Cattaneo, G. Felder,
Comm. Math. Phys. 212 (2000) 212; Mod. Phys. Lett. A16 (2001)
179; Lett. Math. Phys. 56, (2001) 163 .
\item M. Alexandrov, M. Kontsevich, A. Schwarz and O. Zaboronsky, Int. J.
Mod. Phys. A12 (1997) 1405.
\item I. Batalin,  R. Marnelius,
 Phys. Lett. B512 (2001) 225;
Superfield algorithm for topological field theories, in: Multiple
Facets of Quantization and Supersymmetry, M. Marinov Memorial
Volume, ed. M. Olshanetsky, A. Vainshtein,\nolinebreak{\,
}WSPC\nolinebreak{\,}(2002)\nolinebreak{\,}233.
\item A.A. Reshetnyak, General Superfield Quantization Method. I.
Lagrangian Formalism of $\theta$-Superfield Theory of Fields,
ArXiv:hep-th/0210207; Subm. to Int. J. Mod. Phys. A (2003).
\item A.A. Reshetnyak, General Superfield Quantization Method. II.
Hamiltonian Formalism of $\theta$-Superfield Theory of Fields,
ArXiv:hep-th/0303262.
\item A.A. Reshetnyak, General Superfield Quantization Method. III.
Construction of Quantization Scheme, ArXiv:hep-th/0304142.
\end{enumerate}
\end{document}